\documentclass[12pt,preprint]{aastex}
\usepackage{emulateapj5}
\usepackage{apjfonts}
\usepackage{epsfig}

\submitted{Accepted for Publication in ApJ Letters}

\shorttitle{High Levels of Circular Polarization in 3C\,84}
\shortauthors{Homan and Wardle}

\begin{document}

\title{High Levels of Circularly Polarized Emission from the Radio Jet
in NGC 1275 (3C\,84)}
\author{D. C. Homan\altaffilmark{1,2} and J. F. C. Wardle\altaffilmark{3}}

\altaffiltext{1}{Karl Jansky Fellow, National Radio Astronomy Observatory,
Charlottesville, VA 22903}
\altaffiltext{2}{Department of Physics and Astronomy, Denison University,
Granville, OH 43023; homand@denison.edu}
\altaffiltext{3}{Department of Physics, Brandeis University,
Waltham, MA; wardle@brandeis.edu}

\begin{abstract}
We present multi-frequency, high resolution VLBA circular polarization 
images of the radio source 3C\,84 in the center of NGC 1275.  Our images 
reveal a complex distribution of circular polarization in the inner 
parsec of the
radio jet, with local levels exceeding $3$\% polarization, the highest
yet detected with VLBI techniques.  The circular polarization changes
sign along the jet, making 3C\,84 also the first radio jet to show
both signs of circular polarization simultaneously.  The spectrum
and changing sign of the circular polarization indicate that it is 
unlikely to be purely intrinsic to the emitted synchrotron radiation.  
The Faraday conversion process makes a significant and perhaps 
dominant contribution to the circular polarization, and the observed
spectrum suggests the conversion process is near saturation.  The sign 
change in the circular polarization along the jet may result from 
this saturation or may be due to a change in magnetic field order after 
an apparent bend in the jet.  From the small spatial scales probed here, 
$\simeq 0.15$ pc, and the comparably high levels of circular polarization 
inferred for the intra-day variable source PKS 1519$-$273, we suggest a 
connection between small spatial scales and efficient production of circular
polarization.
\end{abstract}


\section{Introduction}
\label{s:intro}
Circular polarization from synchrotron emitting radio jets 
can be a powerful diagnostic of the 
relativistic fluid transported by the jets.  In principle,
circular polarization (CP) observations can be used in concert
with linear polarization and spectral information to constrain
the low energy end of the relativistic particle distribution,
measure magnetic field order and strength, and perhaps even deduce
the particle content, $e^+e^-$ vs. $p^+e^-$, of jets \citep{WHOR98}.

For inhomogeneous synchrotron 
radiation, circular polarization may be produced as an intrinsic part 
of the emitted radiation or through the process of Faraday conversion of 
linear polarization to circular \citep{JOD77}.  On theoretical 
grounds \citep[e.g.][]{Jones88} the Faraday conversion process 
is expected to dominate over the intrinsic CP of 
synchrotron radiation in typical jets.

In recent years circular polarization has been detected 
in a wide range of synchrotron sources: a large number of powerful 
active galaxies \citep*{WHOR98,HW99,RNS00,HAW01}, low-luminosity AGN such
as Sagittarius A* and M81* \citep*{BFB99,SM99,BBFM01,BFM02}, intra-day
variable sources \citep{MKRJ00}, and galactic micro-quasars
\citep{Fend00,Fend02}.  However, the spectral evidence from those
objects with constraints at multiple frequencies creates a 
confusing picture which could be consistent with either intrinsic 
CP or Faraday conversion \citep{BFB99,SM99,MKRJ00,Fend00,BBFM01,Fend02}. 
Only in the case of 3C\,279 \citep{WHOR98} is the Faraday conversion 
process strongly favored by direct observational evidence.  

Here we present multi-frequency Very Long Baseline 
Array\footnote{The VLBA is operated by the National Radio Astronomy 
Observatory which is a facility of the National Science Foundation 
operated under cooperative agreement by Associated Universities, Inc.} 
(VLBA) circular polarization images of the radio 
source 3C\,84 in the center of NGC 1275 ($z = 0.017$).  Because of its
proximity, we have very high spatial resolution, $\simeq 0.15$ pc, in
the radio jet of 3C\,84.  We report very strong circularly polarized emission
in an optically thin region, approximately $0.5$ pc south of the peak radio 
emission.  In \S{\ref{s:obs}} we describe our observations and calibration 
techniques.  Our results are presented in \S{\ref{s:res}}, and we 
analyze and discuss them in \S{\ref{s:dis}}.

\section{Observations}
\label{s:obs}

We observed 3C\,84 in December of 1997 (epoch
1997.94) with the VLBA at four frequencies: 5, 8, 15, and 22 GHz.  
Our nearly full track observations were divided between the
observing bands, giving excellent (u,v)-plane coverage and 
approximately 2 hours on-source per frequency.  Here we present
and discuss the CP results from our 15 and 22 GHz bands.  
Total intensity (Stokes $I$) images are presented in figure 1.

To calibrate the antenna gains to detect circular polarization, 
we used a modification of the {\em Zero-V} self-calibration technique 
described by \citet{HW99}.  {\em Zero-V} self-calibration 
assumes there is no real circular polarization in the data and
calibrates the baseline correlations ($RR = I + V$, $LL = I - V$) versus
a pure Stokes $I$ model.  The derived antenna gains will act to 
suppress any real Stokes $V$ in the data; however, except in the
case of a point source, the gain ``corrections'' cannot remove 
the CP signal completely because the antenna gains are 
multiplicative while $V$ is additive in the baseline 
correlations.  

For an extended source, like 3C\,84, the effect of {\em Zero-V} 
self-calibration on real Stokes $V$ will be to reduce
its level and induce $V$ of the opposite sign on other strong
structure in the source.  It is possible to reconstruct the
original circular polarization distribution through careful modeling
procedures, and that is what we have done.  For the 15 and 22 GHz
results presented here, the core region is well resolved with a 
slowly 
\begin{center}
\figurenum{1}
\epsfig{file=f1.eps,width=2.0in}
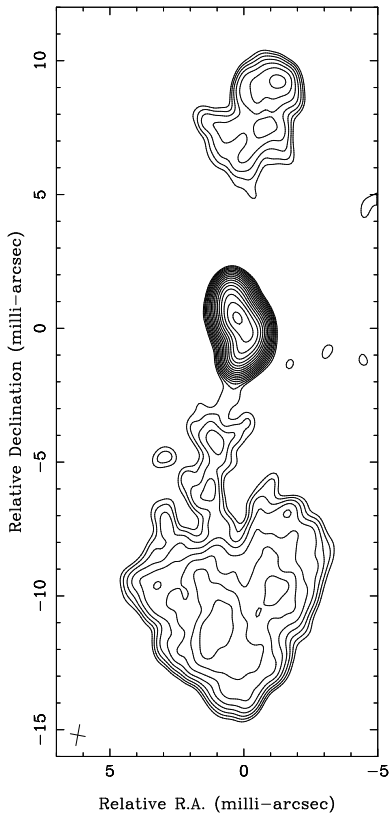
\figcaption{\label{f:i}VLBA Stokes $I$ images of 
3C\,84 at 15 GHz.  The
base contour is 10 mJy/beam, and the contours increase in steps of
$\sqrt{2}$.  The peaks intensity in the image is
2.79 Jy/beam.  The FWHM dimensions of the naturally weighted 
beam ($0.57\times0.80$ mas$\times$mas) is depicted by a cross 
figure in the lower left-hand corner of 
each panel.}
\end{center}
declining brightness profile, so {\em Zero-V} self-calibration
will not greatly distort the real Stokes $V$ \citep{HW99}; this fact 
makes the reconstruction process simple and robust.

Figure 2 shows
the core region of 3C\,84, the distorted CP-signal from {\em Zero-V}
self-calibration, and the reconstruction of the true CP distribution.
The reconstruction was done by assuming the peak CP in the distorted 
map was real, subtracting this CP peak from the original data (prior to 
self-calibration assuming $V = 0$), calibrating the new subtracted
data-set assuming $V=0$, and examining the resulting residual map.  
After one iteration, the residual $V$ map still showed signs of 
real CP which had been distorted by assuming $V=0$. (We expect 
this because the original subtraction could not have been large enough.  
The peak that was subtracted had already been suppressed by some factor
in the initial calibration assuming $V=0$.)  So this process was 
repeated through a number of iterations until the final residual map
was flat.  We then restored all the subtracted components to the 
calibrated data.  

We ran a large number of tests on this CLEAN-type 
algorithm\footnote{Our procedure differs from
a traditional CLEAN-type algorithm in that we improve the 
calibration with each cycle.  We found a loop gain of unity 
to work well in these cases and tests with a loop gain of one-half 
showed no significant difference in the results.} using simulated
data, and we found the reconstruction to be robust, 
particularly for the bright positive components in figure 2.  The 
negative features are real; however, our tests on simulated data 
showed that the reconstruction algorithm typically reduces their amplitude by 
$\sim10$\% at 15 GHz and $\sim30-40$\% at 22 GHz and may introduce 
small shifts in the their peak position of $\sim 0.1$ mas.  
These effects are both due to the weaker nature of negatively polarized
components and their closer proximity to the peak of the total 
intensity structure where the most distortion will occur from our
initial assumption of $V=0$.

\section{Results}
\label{s:res}

Our circular polarization images of 3C\,84 are displayed in figures 2 and 3,
and we find very strong CP at both 15 and 22 GHz.   The registration between
the two frequencies was performed on the Stokes $I$ images alone and is illustrated
for Stokes $V$ in figure 3.  Both the positive and the negative features appear 
at the same locations at the two frequencies frequencies to within a small 
fraction of a beam-width.  All the circularly polarized emission is located 
south of the map peak, with the change in sign from positive to negative 
occurring after the jet appears to bend to the east.  

The bright knot of positive circular polarization (component ``A'' in figure 3) 
is unresolved at 15 GHz and only barely resolved along the jet at 22 GHz.  It
is located in a part of the jet that has a
smoothly declining brightness profile with no distinct features 
in total intensity.  This region of the jet is optically thin 
with a spectral index of $\alpha = -1$ to $-1.5$ ($S\propto\nu^{+\alpha}$) 
for Stokes $I$ between 15 and 22 GHz.  
The fractional local circular polarization for this spot is 
$m_c = +3.2\pm0.1$\% at 15 GHz and $m_c=+2.3\pm0.2$\% at 22 GHz.  
The values are taken at the location of the Stokes $V$ 
peak, and we can find the spectral index of the fractional
circular polarization: $m_c \propto \nu^{-0.9\pm0.3}$.  Note that 
the data were weighted and tapered at the two frequencies to give 
an effectively matched resolution for these measurements.

\begin{figure*}[t]
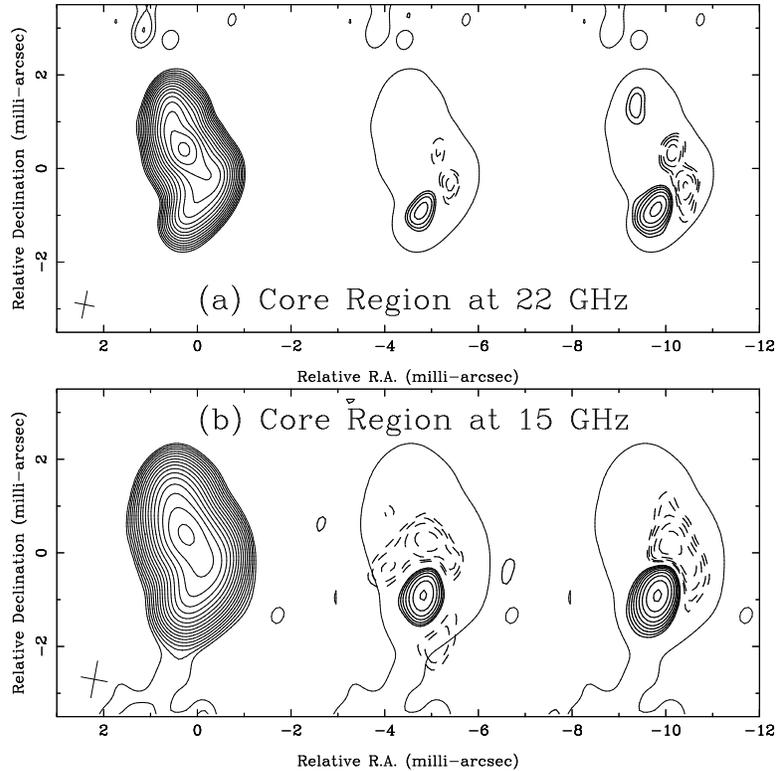

\begin{center}
\epsfig{file=f2a.eps,angle=-90,width=4.0in}\\
\epsfig{file=f2b.eps,angle=-90,width=4.0in}
\end{center}
\figcaption{\label{f:cp}Core region of 3C\,84 at 22 GHz (a) and 
15 GHz (b).
{\em Left:} Stokes $I$ with contours beginning at 10 mJy/beam
and increasing in steps of $\sqrt{2}$.  {\em Center:} Stokes $V$,
calibrated assuming $V=0$, creating some distortion to the
true $V$ distribution.  {\em Right:}  Stokes $V$, reconstruction
of the true distribution using the algorithm described in $\S{3.1}$.
The $V$ peaks in the reconstructed images are 12.7 mJy/beam at 22 GHz
and 33.8 mJy/beam at 15 GHz.  For all Stokes-$V$ images the contours 
begin at 2 mJy/beam and increase in steps of $\sqrt{2}$.  A single 
Stokes $I$ contour is placed around the $V$ images to show 
registration. The FWHM dimensions of the naturally weighted beams 
(($0.42\times0.59$ mas$\times$mas at 22 GHz; $0.57\times0.80$ mas$\times$mas at 15 GHz)
are depicted by a cross figure in the lower left-hand corner of each panel.}   
\end{figure*}

The negatively polarized components (``B'' and ``C'' in figure 3)
are further upstream where the jet is optically thick with 
$\alpha \simeq +1.2$ at the location of ``B'' and 
$\alpha \simeq +1.5$ at the location of ``C''.  These components have 
fractional circular polarizations
of $m_c=-0.7\pm0.1$\% at 15 GHz and $m_c = -1.3\pm0.2$\% at 22 GHz for 
component ``B'' and  $m_c=-0.6\pm0.1$\% at 15 GHz and $m_c = -1.0\pm0.2$\% 
at 22 GHz for component ``C''.   These values include a $+10$\%
amplitude correction at 15 GHz and a $+30$\% amplitude correction
at 22 GHz to compensate for the typical reductions introduced by
our calibration procedure, see \S{2}.  The spectral indicies of
fractional polarization for these negative components are 
$m_c \propto \nu^{+1.7\pm0.6}$ for ``B'' and
$m_c \propto \nu^{+1.4\pm0.7}$ for ``C''.  

It is important to note that our 5 and 8 GHz images (Homan and Wardle, 
in prep.)  show no significant circular polarization in this 
core region, although a predominantly negative sign of CP is seen 
farther south in the jet at 5 GHz.  The core region is poorly resolved
at these longer wavelengths and has become optically thick, so both
blending and optical depth may play a role in the lack of detected CP.  
At the approximate location of component ``A'', the Stokes-I spectrum is flat 
between 8 and 15 GHz and becomes inverted between 5 and 8 GHz, 
suggesting that the self-absorption turnover for this part of the jet
occurs somewhere between 8 and 15 GHz.  Assuming $\tau = 1$ at 10 GHz, we 
estimate the optical depths at 15 GHz and 22 GHz to be $\tau_{15} \sim 0.2$ and 
$\tau_{22} \sim 0.05$ respectively for component ``A''.  

No significant linear polarization is detected in the core 
region at any of our frequencies with an approximate upper 
limit of $1$\% at 22 GHz at the location of the strong, 
positive CP signal analyzed here.  The lack of significant linear
polarization is most likely explained by external 
depolarization in a thermal Faraday screen \citep{W71} 
which will have a negligible effect on the 
circular polarization \citep{JOD77,HAW01}.

\begin{center}
\figurenum{3}
\epsfig{file=f3.eps,angle=-90,width=3.2in}
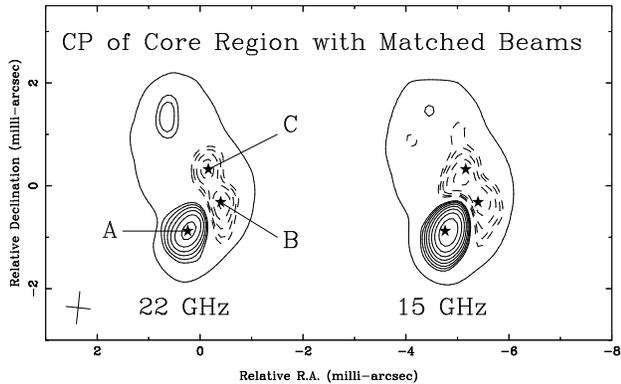
\figcaption{\label{f:cpmatch} Circular polarization of core region
of 3C\,84 with matched beams at 15 and 22 GHz.  The positions of three
circularly polarized components are marked in the 22 GHz image.  These 
same positions are indicated in the 15 GHz image to show registration.  
A single 
Stokes $I$ contour is placed around the $V$ images to show 
registration.
The common FWHM restoring beam 
($0.47\times0.63$ mas$\times$mas) is 
depicted by a cross figure in the lower left-hand corner of the image.
}
\end{center}

\section{Discussion and Conclusions}
\label{s:dis}

We first consider the spot of positive circular polarization 
which is extremely strong at $+3$\% of the {\em local} Stokes $I$ at 
15 GHz.  If the circular polarization is intrinsic to the synchrotron 
radiation itself, its fractional level varies as $m_c \propto \nu^{-0.5}$
for optically thin emission.  This is flatter than our observed spectrum 
but differs from it by only $1.5\sigma$.  The full expression is 
\citep{JOD77,WH03} 

\begin{equation}
m_c = \epsilon_\alpha^v(\nu_{B_\perp}/\nu)^{0.5}\frac{B_{uu}\cos\theta}{B_\perp^{\,rms}}
\end{equation}

\noindent where $\epsilon_\alpha^v$ is a constant $\approx 2$ for 
$\alpha < -1$
\citep{JOD77} and $\nu_{B_\perp} = 2.8B_\perp^{\,rms}$ MHz is the 
gyro-frequency for $B_\perp^{\,rms}$ in Gauss.  $B_{uu}\cos\theta$ 
is the component of uniform, unidirectional magnetic field along the 
line of sight responsible for generating the circular polarization.  
$B_\perp^{\,rms}$ is the RMS averaged component of field in the plane 
of the sky and includes contributions not only from the transverse
part of $B_{uu}$ but also from disordered field or ordered loops of 
field which will contribute to Stokes $I$ but not $V$.  

If the circular polarization is entirely due to the intrinsic
mechanism, we can estimate the minimum magnetic field strength
required by the very high levels of circular polarization 
detected at 15 GHz.  \citet*{WRB94} find the jet of 3C\,84
to be only mildly relativistic and oriented at a large angle
to the line of sight, $\theta = 30^\circ-55^\circ$.  We take 
the Doppler factor to be unity and $\theta = 30^\circ$ in these
calculations. The largest possible circular polarization will
be produced by the intrinsic mechanism when the magnetic field is 
completely uniform and uni-directional.  Under these conditions,
$B_\perp^{\,rms} = B_{uu}\sin\theta$, and we find that a uniform field
strength of $B_{uu} \simeq 1$ Gauss is necessary to produce the observed 
$3.2$\% circular polarization at $\nu = 15$ GHz.  Any sign reversals
or disorder in the magnetic field will increase the required
magnetic field strength.

Given the above estimate, it is possible that a very strong ($\sim 1$ G), 
uniform magnetic field along the jet axis is responsible for generating
the observed positive circular polarization via the intrinsic mechanism alone.  
However, the negative components of circular polarization cannot
be easily explained by the intrinsic mechanism.  They are located upstream 
where the jet is optically thick.   \citet{PS71}\footnote{The plots given by
\citet{JOD77} consider the combined effects of intrinsic circular polarization
and Faraday conversion and here we are interested in the intrinsic mechanism
alone.  We verified the \citet{PS71} result by direct numerical integration
of the Jones and O'Dell radiative transfer formulas.} showed that intrinsic circular 
polarization will change sign when the optical depth is significantly greater than 
unity, but their results
also show that the spectral index of the fractional polarization, $m_c$, can only
be negative after the sign change.   This result conflicts with our observations
which show a positive spectral index for $m_c$ for both negatively polarized 
components.  

An alternative to a pure intrinsic circular polarization model is to
have a significant (and perhaps dominant) contribution from the
Faraday conversion effect.
High levels of circular polarization can be easily generated via
the Faraday conversion process which converts linear polarization
to circular \citep{Jones88}.  Faraday conversion 
operates on Stokes $U$ relative to the local magnetic field 
orientation (while synchrotron radiation from the same field
generates only Stokes $Q$) and therefore requires 
either some change in field orientation along the line of sight or 
some internal Faraday rotation to drive the process.  If the 
conversion process is 
driven by Faraday rotation, some uniform, unidirectional
magnetic field, $B_{uu}$, is required to generate internal Faraday
rotation, but here we can have $B_{uu}\ll B_\perp^{\,rms}$.  

In the optically thin regime, the conversion process 
has a steep spectrum, $m_c \propto \nu^{-3}$, or even steeper if
it is driven by Faraday rotation \citep{WH03}.  
For the positively polarized, optically thin component, our 
observed spectrum is flatter than this, $\nu^{-0.9\pm0.3}$, and would seem 
to reject Faraday conversion as the dominant process.  However, at 
$\tau_{15}\sim0.2$, we are not completely
optically thin, and the predicted spectrum flattens as 
we approach $\tau = 1$ \citep{JOD77}.  
This is due to optical depth and saturation in the conversion process 
as it cycles into a regime where it generates the opposite sign of circular 
polarization, depolarizing the radiation and flattening the spectrum.  
This saturation most easily occurs when the conversion process is driven 
by internal Faraday rotation, as the Faraday rotation will also saturate and 
begin to depolarize the linear polarization, reducing the amount available
for conversion and also flattening the spectrum.

Detailed modeling of this saturation effect is beyond the scope of 
this paper; however, it is possible to explain not only
the spectrum of the positively polarized component but also the
change in sign and the spectrum of the negatively polarized components
via this mechanism.  Of course, the sign change of polarization 
from negative to positive may not be due to saturation but 
rather due to changes in magnetic field order.  If the Faraday
conversion is driven by Faraday rotation, it is 
possible to flip the net field polarity and generate the opposite 
sign of circular polarization.  
If the Faraday conversion is not driven by Faraday rotation but 
rather results from a changing field order along the line of 
sight, as might be the case for a helical field, a change in that 
field order, such as the pitch angle of the helix, could also 
change the sign of the resulting circular polarization.  

Another interesting aspect of the circular polarization distribution
is its apparent break-up into three distinct circularly polarized
components while the total intensity (Stokes-$I$) distribution
appears smooth and continuous throughout the region.  The sign 
change of circular
polarization between components A and B will naturally make those
regions appear separated, since Stokes-$V$ must pass through
zero between them.  The discrete nature of 
component C is more difficult to understand.  We note that it is closer
to the base of the jet and offset from its center, so its discrete 
appearance may be due
to a fortuitous location where the optical depth is not too large
and other parameters are suitable for producing significant CP. 
The magnetic field order in the jet and the presence (or absence) of
shocks in the magnetic field may also create enhanced regions of 
circular polarization; however, without linear polarization information
it is impossible to evaluate those possibilities.

Finally, we note that 3C\,84, at a redshift of 
$z=0.017$, shows the strongest fractional circular polarization yet 
detected by the VLBA.  Much more powerful quasars rarely show local 
CP stronger than $0.3$\% \citep{HW99,HAW01}; however, these sources 
are much farther away and at such distances 3C\,84 would have only 
$0.1-0.2$\% local circular polarization 
due to blending with the core.  In this light, it is interesting that
the intra-day variable source PKS 1519$-$273 exhibits
CP at the $2-4$\% level in a scintillating 
component $15-35$ $\mu$as in size \citep{MKRJ00}.  This 
corresponds to a region of $0.05-0.30$ pc in size (for $z\gtrsim0.2$,
\citealt{S93}), 
comparable to the linear resolution we achieve in these
observations, $\simeq 0.15$ pc. The coincidence 
of high levels of CP with small spatial 
scales may be due to to the increased relative importance of 
uni-directional magnetic field ($B_{uu}$) in the jet at small 
radii \citep{WH03}.

\acknowledgments

This work has been supported by the National Radio Astronomy
Observatory and by NSF grant AST 99-00723.


\begin{thebibliography}
\expandafter\ifx\csname natexlab\endcsname\relax\def\natexlab#1{#1}\fi


\bibitem[{{Bower} {et~al.}(1999){Bower}, {Falcke}, \& {Backer}}]{BFB99}
{Bower}, G.~C., {Falcke}, H., \& {Backer}, D.~C. 1999, \apjl, 523, L29

\bibitem[{{Bower} {et~al.}(2002){Bower}, {Falcke}, \& {Mellon}}]{BFM02}
{Bower}, G.~C., {Falcke}, H., \& {Mellon}, R.~R. 2002, \apjl, 578, L103

\bibitem[{{Brunthaler} {et~al.}(2001)}]{BBFM01}
{Brunthaler}, A., {Bower}, G.~C., {Falcke}, H., \& {Mellon}, R.~R. 2001, \apjl,
  560, L123

\bibitem[{{Fender} {et~al.}(2000){Fender}, {Rayner}, {Norris}, {Sault}, \&
  {Pooley}}]{Fend00}
{Fender}, R., {Rayner}, D., {Norris}, R., {Sault}, R.~J., \& {Pooley}, G. 2000,
  \apjl, 530, L29

\bibitem[{{Fender} {et~al.}(2002){Fender}, {Rayner}, {McCormick}, {Muxlow},
  {Pooley}, {Sault}, \& {Spencer}}]{Fend02}
{Fender}, R.~P., {Rayner}, D., {McCormick}, D.~G., {Muxlow}, T.~W.~B.,
  {Pooley}, G.~G., {Sault}, R.~J., \& {Spencer}, R.~E. 2002, \mnras, 336, 39

\bibitem[{{Homan} {et~al.}(2001){Homan}, {Attridge}, \& {Wardle}}]{HAW01}
{Homan}, D.~C., {Attridge}, J.~M., \& {Wardle}, J.~F.~C. 2001, \apj, 556, 113

\bibitem[{{Homan} \& {Wardle}(1999)}]{HW99}
{Homan}, D.~C. \& {Wardle}, J.~F.~C. 1999, \aj, 118, 1942

\bibitem[Homan \& Wardle(2000)]{HW00} Homan, D.~C.~\& 
Wardle, J.~F.~C.\ 2000, \apj, 535, 575

\bibitem[Homan \& Wardle(2003)]{HW03}
Homan, D.~C., \& Wardle J.~F.~C. 2003, In Circular Polarization From Relativistic
Jet Sources, a workshop held in Amsterdam, NL, 17-19 July, 2002, eds. R.~P. Fender
\& J.-P. Macquart, \apss, 288, 29

\bibitem[{{Jones}(1988)}]{Jones88}
{Jones}, T.~W. 1988, \apj, 332, 678

\bibitem[{{Jones} \& {O'Dell}(1977)}]{JOD77}
{Jones}, T.~W. \& {O'Dell}, S.~L. 1977, \apj, 214, 522

\bibitem[{{Macquart} {et~al.}(2000){Macquart}, {Kedziora-Chudczer}, {Rayner},
  \& {Jauncey}}]{MKRJ00}
{Macquart}, J.-P., {Kedziora-Chudczer}, L., {Rayner}, D.~P., \& {Jauncey},
  D.~L. 2000, \apj, 538, 623

\bibitem[Marscher(1987)]{M87} Marscher, A.~P.\ 1987, In 
Superluminal Radio Sources, eds. J.~A. Zensus \& T.~J. Pearson 
(Cambridge: Cambridge Univ. Press), 280 

\bibitem[Pacholczyk \& Swihart(1971)]{PS71} 
Pacholczyk, A.~G.~\& Swihart, T.~L.\ 1971, \apj, 170, 405

\bibitem[{{Rayner} {et~al.}(2000){Rayner}, {Norris}, \& {Sault}}]{RNS00}
{Rayner}, D.~P., {Norris}, R.~P., \& {Sault}, R.~J. 2000, \mnras, 319, 484


\bibitem[{{Sault} \& {Macquart}(1999)}]{SM99}
{Sault}, R.~J. \& {Macquart}, J.-P. 1999, \apjl, 526, L85

\bibitem[Stickel, Fried, \& Kuehr(1993)]{S93} Stickel, M., 
Fried, J.~W., \& Kuehr, H.\ 1993, \aaps, 98, 393

\bibitem[Walker et al.(1994)Walker, Romney, \& Benson]{WRB94} Walker, 
R.~C., Romney, J.~D., \& Benson, J.~M.\ 1994, \apjl, 430, L45

\bibitem[{{Wardle}(1971)}]{W71}
{Wardle}, J.~F.~C. 1971, \aplett, 8, 183

\bibitem[Wardle \& Homan(2003)]{WH03}
Wardle, J.~F.~C., \& Homan, D.~C. 2003, In Circular Polarization From Relativistic
Jet Sources, a workshop held in Amsterdam, NL, 17-19 July, 2002, eds. R.~P. Fender
\& J.-P. Macquart, \apss, 288, 143

\bibitem[{{Wardle} {et~al.}(1998)}]{WHOR98}
{Wardle}, J.~F.~C., {Homan}, D.~C., {Ojha}, R., \& {Roberts}, D.~H. 1998, \nat,
  395, 457


\end{thebibliography}
\end{document}